\documentclass{appolb}
\usepackage{epsfig,amssymb,amsmath}
\def\bge{\begin{equation}}
\def\ene{\end{equation}}
\def\bg{\begin{eqnarray}}
\def\en{\end{eqnarray}}

\def\e{\epsilon}
\def\bge{\begin{equation}}
\def\ene{\end{equation}}
\def\bg{\begin{eqnarray}}
\def\en{\end{eqnarray}}

\def\e{\epsilon}

\begin{document}
% \eqsec  % uncomment this line to get equations numbered by (sec.num)
\title{
%
%\hbox{
Gluons in the $\eta'$ and in nucleon resonances
\thanks{Presented at the Workshop at 1 GeV scale: from mesons to axions, Krakow, September 19-20 2024.}
}
\author{Steven D. Bass 
\address{
Marian Smoluchowski Institute of Physics, 
Jagiellonian University, \\ 
Krakow, Poland \\
\vspace{3ex}
Kitzb\"uhel Centre for Physics, Kitzb\"uhel, Austria}
}
\maketitle
\begin{abstract}
\noindent
We discuss the role of gluon dynamics in  $\eta'$ physics and in nucleon resonances where excitations of gluonic potentials 
may also be important. 
Interesting phenomenology includes a possible narrow near threshold resonance in $\eta'$ photoproduction and 
whether the parity doublets observed in the higher mass 
nucleon resonance spectrum
might be hinting at a 
possible second minimum in the confinement potential 
corresponding to 
supercritical confinement.
\end{abstract}

\section{Introduction}

QCD is built from quark and gluon interactions with the gauge symmetry of SU(3)$_c$.
QCD is characterised by asymptotic freedom in the ultraviolet driven by the 3 gluon vertex with gluons carrying colour charge as well as quarks and by confinement and dynamical chiral symmetry breaking in the infrared. 
Effects of the 3 gluon vertex are also seen in gluon jets produced at high energy colliders and in the QCD evolution of parton distribution functions.
The non abelian gauge symmetry is also characterised by 
non-perturbative gluon topology 
-- a property of 
gluonic degrees of freedom 
that is insensitive to local deformations of the gluon fields. 
This gluon topology plays an essential role in the generating the large mass of the $\eta'$ meson. It may also play a vital role in QCD quark and gluon confinement.
Hadrons -- the ground state excitations of the theory -- are colourless baryons and mesons with the quarks and antiquarks confined in some gluon induced potential.
Non-perturbative glue also drives the asymptotic behaviour of high energy hadron scattering processes via so called pomeron Regge exchanges 
with connection to possible new glueball states.
In this contribution we focus on the role of gluonic degrees of freedom in the $\eta'$--proton and $\eta'$--nucleus systems at low energies and in the gluonic potential that determines the nucleon resonance spectrum. 
We discuss the phenomenology of exciting the 
non-perturbative glue associated with the large $\eta'$ mass in photoproduction reactions. This glue must be active in the pion cloud of the nucleon to suppress any isoscalar pion degrees of freedom.
We also make some remarks on possible new structure in the confining potential. In particular, we discuss whether parity doublets observed in the higher mass N* and 
$\Delta$* resonances, starting 
$\approx$ 700 MeV above the proton and $\Delta$(1232) masses, 
might be hinting at a second minimum in the confinement potential with the confinement dynamics working differently to that associated with the proton bound state.

\section{Gluon topology and DChSB}

Low energy QCD is characterised by confinement and dynamical chiral symmetry breaking, DChSB. The chiral symmetry between 
left-handed and right-handed quarks is spontaneously broken by the scalar quark condensate in the vacuum. 
One finds 
Nambu–-Goldstone bosons
which satisfy the Gell-Mann–-Oakes–-Renner 
relation
\begin{equation}
m_P^2 
f_{\pi}^2 = - m_q \ \langle {\rm vac} | \ {\bar \psi} \psi \ | {\rm vac} \rangle 
+ {\cal O} (m_q^2)
\end{equation}
with $f_{\pi} = \sqrt{2} F_{\pi} = 131$ MeV.
The mass squared of the Goldstone bosons $m_P^2$
is at leading order
proportional to the mass of their valence quarks. 
This works well for the pions and kaons. 
The lightest mass pions have mass
135 MeV for the $\pi^0$ 
and 140 MeV for the charged $\pi^{\pm}$.  
The kaon masses are
$m_{K^\pm} =  494$ MeV and
 $m_{K^0} = m_{{\overline K}^0} = 498$ MeV.
This picture when taken alone comes with the issue
that it gives 9=8+1 
Nambu–-Goldstone bosons in total, 8 from SU(3) and one from axial U(1). 
Treating the isoscalar partners only in the spirit of Eq.~(1) with $m_P^2 \propto m_q$
the $\eta'$ would come out as a strange quark state with mass 
$\sqrt{2 m_K^2 - m_\pi^2}$ 
and the $\eta$ would be a light quark state degenerate with the pion 
after $\eta$--$\eta'$ mixing induced by the strange quark mass. 
The values of the physical meson masses are $m_\eta = 547$ MeV and $m_{\eta '} = 958$ MeV. 
This situation is resolved by gluonic degrees of freedom in the flavour singlet channel.
Working at leading order in the chiral expansion 
if we add a phenomenological singlet glue term, then diagonalising the mass matrix gives
$\eta$ and $\eta'$
masses satisfying 
the Witten–-Veneziano  mass formula 
\begin{equation}
   m_\eta^2 + m_{\eta'}^2 = 2m_K^2 + {\tilde m}_{\eta_0}^2 .
\end{equation}
Within this approximation, 
if we take
${\tilde m}_{\eta_0}^2 =
0.73$ GeV$^2$  for the singlet gluonic mass term then the $\eta$ and $\eta'$ masses each come out correct to within 10\% accuracy with the mixing angle
-20 degrees.

The gluonic mass term  ${\tilde m}_{\eta_0}^2$ 
is associated with non-perturbative gluon topology 
and its effect is  incorporated in axial U(1) extended chiral Lagrangians
-- for reviews see~\cite{Shore:1998dm,Shore:2007yn}. 
The theory involves the interface of local anomalous Ward identities and 
non-local topological structure.
The gluonic mass term has a rigorous interpretation 
as the leading term when one makes an expansion in 
$1/N_c$
($N_c$ is the number of colours) 
with 
$\alpha_s N_c$
and the number of light flavours $N_f$ held fixed.
It is related to a quantity $\chi (0)|_{\rm YM}$ called 
the Yang–-Mills topological susceptibility \cite{Witten:1979vv,
Veneziano:1979ec}
\begin{equation}
{\tilde m}_{\eta_0}^2 \bigg|_{\rm LO}
= - {6 \over f_{\pi}^2} \chi (0) \bigg|_{\rm YM} .
\end{equation}
The quantity $\chi(0)|_{\rm YM}$ is defined through  
the two-point function
\begin{equation}
\chi (k^2)|_{\rm YM} = \int d^4 z \ i \ e^{ik.z} \
\langle {\rm vac} | \ T \ Q(z) Q(0) \ | {\rm vac} \rangle 
\big|_{\rm YM} 
\end{equation}
calculated in the pure glue theory without quarks. 
Here 
$Q = \frac{\alpha_s}{4 \pi} G_{\mu \nu} {\tilde G}^{\mu \nu}$ 
is the 
topological charge density which is a total derivative and 
$T$ denotes the time ordered product.
From this $Q(z)$ one finds the 
topological winding number 
\begin{equation}
    n = \int d^4 z \ Q (z) \in {\rm Z \ or \ Q}, , \ \ \ \ \ \frac{\delta}{\delta A_\mu} n [A] = 0 
\end{equation}
which in QCD 
takes quantised values 
(integers Z and perhaps also fractions Q). 
The winding number is a non-perturbative property and
invariant under local deformations of the gluon fields. 
If we assume that the 
topological winding number remains finite independent of the value of $N_c$ 
then
\cite{Witten:1979vv} 
\begin{equation}
{\tilde m}_{\eta_0}^2 
\sim 1 / N_c \ {\rm as} \ N_c \to \infty .
\end{equation}
Computational QCD lattice calculations
performed with $N_c=3$
of 
the pure gluonic term on the right-hand side of Eq.~(3) 
and the meson mass contributions with dynamical quarks
in the Witten–-Veneziano formula, 
Eq.~(2), 
give excellent agreement at the 10\% percent level
\cite{Cichy:2015jra}.  
One finds
$\chi^{1/4} (0)|_{\rm YM} = 185.3 \pm 5.6$ MeV,  
very close to the phenomenological value 180 MeV 
which follows from taking 
${\tilde m}_{\eta_0}^2 = 0.73$GeV$^2$ 
in Eq.~(2).

DChSB including $U_A$(1) degrees of freedom can be included in 
low energy 
effective chiral Lagrangians~\cite{DiVecchia:1980yfw,Witten:1980sp,Leutwyler:1997yr}.
The gluonic mass term is introduced via a potential involving $Q$ which enters with no kinetic energy term.
That is,   
it does not correspond to a physical glueball.
The low energy Lagrangians can be used to calculate a host of low energy reactions involving the $\eta'$ 
including with new 
gluonic intermediate states 
(OZI violating) mediated by couplings involving $Q$ beyond 
the simplest chiral perturbation theory.
The $\eta'$ is the physical state that comes out with OZI violating couplings to other hadrons including through the $\eta'$--nucleon coupling constant $g_{\eta' NN}$ \cite{Bass:1999is}.
When combining with the gluonic $Q$ one also finds a colour-singlet gluonic object $G$ carrying renormalisation scale dependent couplings to hadrons.
The effect can be seen in the singlet version of the Goldberger–-Treiman relation
\cite{Shore:1991dv} 
which has connections to the proton's internal spin structure~\cite{Bass:2024sgg,Bass:2004xa,Aidala:2012mv}.
This gluonic object is not a physical object.
That is, 
it has no physical state and occurs only in exchanges and virtual loop diagrams.
For detailed discussion see~\cite{Shore:1998dm,Shore:2007yn}.

\section{OZI violation in 
$\eta'$-nucleon and $\eta'$-nucleus interactions}

The gluonic potential in $Q$ that generates the $\eta'$'s
gluonic mass term
also contributes to the $\eta'$--proton 
scattering length 
and to $\eta'$ mass modifications in nuclei and possible $\eta'$ bound states 
in nuclei.
Through the flavour-singlet version of the Weinberg–-Tomozawa relation the 
gluonic mass term 
gives a finite value for the real part of the $\eta'$-nucleon scattering length
even in the Born term contribution 
which is non-vanishing in the chiral limit \cite{Bass:2018xmz}.
It is not unreasonable that this result  generalises to a full non-perturbative analysis with (S-wave) resonances close to threshold. 
Meson mass shifts in medium are related to the meson-nucleon scattering length.
This OZI related scattering length leads one to expect a finite $\eta'$ mass shift in medium as a gluon mediated effect.

$\eta'$ photoproduction experiments at ELSA in Bonn 
from Carbon and Niobium targets have revealed an 
$\approx -40$ MeV shift in the $\eta'$ mass at nuclear matter density. 
The measured $\eta'$-nucleus optical potential has real and imaginary parts $V+iW$ with 
\begin{eqnarray}
 V(\rho=\rho_0) &=& -40 \pm 6 \pm 15 \ {\rm MeV} 
\nonumber \\ 
W(\rho=\rho_0) &=& -13 \pm 3 \pm 3 \ {\rm MeV}   
\label{eq13}
\end{eqnarray}
-- see Refs.~\cite{CBELSATAPS:2013waf,Metag:2017yuh,Bass:2021rch}.
The large relative value of 
the real part compared to the imaginary part hints at the existence of possible bound states in nuclei with a vigorous experimental programme at GSI and in Japan to look for them.
Theoretically, 
this mass shift is  catalysed by the gluonic mass term  contribution. 
Without this the $\eta'$ would be a strange quark state with minimal interaction with the  $\sigma$ (correlated two-pion exchange) mean field in the nucleus. 
The mass shift -37 MeV was predicted by 
Quark Meson Coupling model \cite{Bass:2005hn,Bass:2013nya} 
with the light-quark component in the $\eta'$ coupled to this $\sigma$ mean field and taking an $\eta-\eta'$ 
mixing angle of -20 degrees in the leading order one-mixing-angle scheme.
The optical potential 
in Eq.~(7) is 
related to the $\eta'$-proton scattering length with finite real part expected from anomalous glue.
Numerically Eq.~(7) is consistent with the COSY-11 measurement of the 
$\eta'$-proton scattering length \cite{Czerwinski:2014yot}.

\section{Gluons and the nucleon resonance spectrum}

Besides the special role of glue in the $\eta'$, gluonic potentials are also important in the N* resonance spectrum.
The proton and N* resonance spectrum 
are usually understood as following from solving for quark wavefunctions in some gluonic induced confinement potential.
This gluonic potential behaves like a string of glue with string breaking corresponding to the formation of new hadrons.
The N* resonance spectrum corresponds to quark orbital and radial excitations in this potential. 
The colour hyperfine (one gluon exchange, OGE) and pion cloud corrections also play an important role to get the masses correct. 
For example, 
the nucleon-$\Delta (1232)$ mass splitting comes mainly from the OGE potential~\cite{Close:1979bt}.
Confinement is usually understood in terms of a scalar potential connecting 
left-handed and right-handed quarks.
Scalar confinement spontaneously breaks chiral symmetry along with the vacuum quark condensate in Eq.~(1). 
This leads to parity non-doublets in the light baryon spectrum.
For example, 
the lowest mass negative parity would-be partner of the proton is the 
$S_{11}(1535)$ 
which has mass 597 MeV heavier than the proton.
In quark models and in lattice calculations it is understood as a 3 quark state
(1s)(1s)(1p). 
The pion cloud involves just isovector pions.
There is no isoscalar pion state consistent with the gluonic potential linked to 
${\tilde m}_{\eta_0}^2$
being active in the range of the pion cloud.

This is not the complete story. 
Quark model calculations based on the symmetry groups 
${\rm SU(6)} \otimes {\rm O(3)}$
and present QCD lattice calculations do not give a complete description of the nucleon resonance spectrum.
One also observes parity doublets in
the excited N* and $\Delta$* spectra beyond about 700 MeV above the proton and $\Delta$(1232) masses
\cite{Burkert:2022adb,Anisovich:2011ye}. 
Some states predicted by constituent quark models are absent in experimental data -- the so called missing resonance problem with recent discussion in~\cite{Leinweber:2024psf}.
There are also interesting hints 
for a possible 
narrow resonance in $\eta'$ photoproduction just above threshold 
\cite{Anisovich:2018yoo,Tiator:2018heh}.
In rest of this contribution we discuss possible excitation of the gluonic potential term associated with the $\eta'$ mass and a possible extra minimum in the confinement potential accessible at higher energies allowing
for tunneling between vacuum states. 
This new minimum might be associated with supercritical confinement with analytic properties so that the corresponding bound states behave differently in real world Minkowski-space and in Euclidean-space based lattice calculations.

\subsection{Photoproduction: $\gamma p \to \eta' p$ 
and gluon-excitation resonances}

The gluonic potential in $Q$ 
which generates 
the large $\eta'$ mass contribution
${\tilde m}_{\eta_0}^2$
also acts in the virtual pion cloud. 
The absence of isoscalar pions means that the gluonic potential involving $Q$ and $G$ 
must be active here so it should be considered an essential part of the nucleon. 
Can we excite it?
The extended range of the pion cloud corresponds to a Compton wavelength $\lambda \sim 1/m_\pi$ so the relative momentum of any meson-nucleon final state should be about the scale $m_\pi$. 
Excitation of $Q$ in the nucleon wavefunction 
can evolve only 
into the flavour-singlet parts of the $\eta'$ 
or $\eta$ 
in the final state. 
The $Q$ and $G$ fields do not correspond to a physical glueball
and exist only in intermediate states. 
This would lead to an excited
resonance close to the $\eta'$ production threshold
which decays just to $\eta'$--nucleon final states with small phase space 
(plus a small fraction to $\eta$--nucleon). 
If present  
this is expected to be narrow since the $\eta'$ production proceeds through 
gluonic intermediate states with OZI violation. 
Within the framework of the large $N_c$  approximation OZI violating processes occur with result suppressed 
by powers of $1/N_c$ 
\cite{tHooft:1973alw,Witten:1979kh}.

$\eta'$ photoproduction has been studied in experiments at  
ELSA, GRAAL, JLab and MAMI -- see \cite{Bass:2018xmz},
Key resonances are discussed in \cite{Thiel:2022xtb} 
with strong coupling to the 
N*(1895) with $I=\frac{1}{2}$, 
$J^P=\frac{1}{2}^-$.
which has mass 
$1907 \pm 10$ MeV 
and width $100^{+40}_{-10}$ MeV with decay mode to $\eta' p$ of 10-40\%.

Interestingly, 
the near threshold GRAAL 
beam asymmetry measurement \cite{LeviSandri:2014uhc} and 
Mainz A2 measurement of the differential cross section \cite{A2:2017gwp}  from proton targets  has been 
interpreted as possible evidence for a narrow
near threshold nucleon resonance with mass about 1900 MeV and width about 2 MeV.
These claims follow from coupled channels analyses of the $\eta'$ photoproduction data 
\cite{Anisovich:2018yoo,Tiator:2018heh}.
These data are statistics limited.
To be sure one needs new measurements with
finer energy binning and also including new polarisation observables \cite{Tiator:2018heh}.
With this caveat in mind, 
suppose we take these analyses seriously. 
The Bonn group found a N* resonance 
D$_{13}$(1900) 
with mass 
${\cal M}_{\eta' p} = 1900 \pm 1$ MeV and width 
$\Gamma_{\eta' p} < 3$ MeV 
\cite{Anisovich:2018yoo}.
The Mainz group preferred an  S$_{11}$(1900) with 
${\cal M}_{\eta' p} = 1902.6 \pm 1$ MeV and 
$\Gamma_{\eta' p} = 2.1 \pm 0.5$ MeV.
\cite{Tiator:2018heh}.
Note the different resonance quantum numbers in the two analyses.
If accurate, 
one is looking at a N* resonance 
very close to threshold at 
${\cal M}_{\eta' p} = 1896$ MeV and a small width unusual for hadron physics. 
For example, in comparison, the $\Delta$(1232) and
$S_{11}$(1535) come with widths about 117 and 150 MeV respectively.
The $\eta'$ lifetime is a factor of 10 longer than that of the possible $\eta'$-nucleon resonance with the $\eta'$ total width of 0.2 MeV.
The $\eta'$ is essentially stable compared to 
this N* state (if it exists).

For a resonance with mass 1901 MeV decaying to a proton and an $\eta'$ in the 
centre-of-mass frame one finds
\begin{equation}
    W = 1901 = E_p + E_{\eta'} 
    = m_p + m_{\eta'} + 
    \frac{1}{2} {\rm p}^2 \biggl(\frac{1}{m_p} + \frac{1}{m_{\eta'}} \biggr) + ...
\end{equation}
where ${\rm p}$ denotes the magnitude of the 3-momentum of the outgoing proton or $\eta'$.
The value $W=1901$ MeV corresponds to 3-momentum of 69 MeV 
in the centre of mass frame and between 53 and 81 MeV with a 
2 MeV float on the value of $W$.
This 69 MeV value gives a relative 3-momentum of 138 MeV
which is about the pion mass. 
Using standard quantum relations, this momentum 
corresponds to a distance scale about the pion Compton wavelength.
The wavelength here corresponds to the distance scale characterising the pion cloud as measured through e.g. the neutron charge form factor. 
The $\pm 2$ MeV float 
corresponds to a range of wavelengths between 1.21 fm and 1.85 fm.
This possible narrow resonance 
in $\eta'$ photoproduction close to threshold is interesting. 
Perhaps it might 
be associated with excitation of the $U_A(1)$ gluon potential in the pion cloud (?)
Given the quantum numbers of the $Q$ field the simplest interpretation would be a S$_{11}$ state in this scenario.

\subsection{Gribov confinement and parity doublets in the hadron spectrum}

Parity doublets are seen in higher mass light quark nucleon resonances 
\cite{Burkert:2022adb}.
This is in contrast with the lowest mass hadronic states which clearly exhibit the effect of DChSB.
The first parity doublet excited resonance is  $\approx 700$ MeV above the proton mass 
-- see Table I -- likewise also for the $\Delta$. 
These parity doublets are not seen in QCD lattice calculations and are also 
not seen in the strange baryon quark resonances \cite{Jaffe:2006jy}.
They are a puzzle in the usual quark model picture with a scalar confinement potential and single minimum inducing DChSB.
These doublet states have attracted considerable theoretical attention
-- see e.g. 
\cite{Jaffe:2006jy,Glozman:1999tk,Brodsky:2014yha}.

\begin{table}[t!]%[htbp]
\caption{
Parity doublets observed in excited nucleon resonances. 
\label{restable}}
\begin{center}
\begin{tabular}[t]{llll}
\hline
N$_{1/2^+}$(1710) & N$_{1/2 ^-}$(1650) & \ \ \ 
\ \ \
N$_{3/2^+}$(1720) &
N$_{3/2 ^-}$(1700) \\
N$_{5/2^+}$(1680) & N$_{5/2^-}$(1675) &
\ \ \ 
\ \ \
N$_{1/2^+}$(1880) & N$_{1/2^-}$(1895) \\
N$_{3/2^+}$(1910) &
N$_{3/2^-}$(1875) &
\ \ \ 
\ \ \
N$_{5/2^+}$(2095) & N$_{5/2^-}$(2075) \\
N$_{7/2^+}$(2100) & N$_{7/2^-}$(2190) &
\ \ \ 
\ \ \
N$_{9/2^+}$(2220) & N$_{9/2^-}$(2250) \\
\hline
$\Delta_{1/2^+}$(1910)
& $\Delta_{1/2^-}$(1900)
& 
\ \ \ 
\ \ \
$\Delta_{3/2^+}$(1920)
& $\Delta_{3/2^-}$(1940) \\ 
$\Delta_{5/2^+}$(1905) 
& 
$\Delta_{5/2^-}$(1930) 
&
\ \ \ 
\ \ \
$\Delta_{7/2^+}$(1950)
&
$\Delta_{7/2^-}$(2200) (?) \\
\hline
\end{tabular}
\end{center}
\end{table}

Looking for clues,
it is interesting to also consider the possibility of an extra minimum in the confinement potential for the light up and down quarks. 
To avoid clash with successful phenomenology, 
this second minimum 
should lie 
$\approx 700$ MeV above the potential minimum corresponding 
to the proton bound state.
The quarks could then  tunnel 
to a new confining vacuum state once the excited resonance energy reaches the energy difference between the two minima.
\footnote{A possible second minimum in the electroweak Higgs potential is discussed in connection with stability of the Higgs vacuum for the Standard Model;   for details see~\cite{Bass:2021acr} and references therein.}
With parity doublets in mind 
suppose that the confining solution
in the second minimum
corresponds to a vector interaction,
viz. 
instead of the usual scalar confinement potential that induces mass splitting between positive and negative parity states. 
Further still, suppose that the 
singularity structure of the new confining quark propagator here 
does not allow Wick rotation through 
analytic continuation between Minkowski and Euclidean space.
In this case the theories in Euclidean and Minkowski space would be different. 
Lattice calculations are performed in Euclidean space  
so would miss the states defined with respect to the second minimum. 
Qualitatively, 
this scenario would seem to have the ingredients needed 
to describe the parity doublet states.

Interestingly, 
these properties are 
characteristics of Gribov's supercritical confinement scenario \cite{Gribov:1986vp,Gribov:1999ui}.
With supercritical confinement
the attractive 
colour Coulomb interaction with large coupling $\alpha_s$, 
drags the dynamical quark below the Fermi surface of the Dirac sea. 
What follows is a rearrangement of vacuum energy levels with the dynamical quark tumbling through the vacuum -- a phenomena called ``falling into the centre''. 
Both positive and negative energy states become unstable.
The resulting confining quark 
propagator has analytic structure including cuts that do not permit a  straightforward 
continuation 
between Minkowski and Euclidean space.
That is, the Wick rotation implicit in lattice calculations would not be working for these states.
Physically  
one might think of the dynamical quark colour charge as falling through the continuum of negative energy levels until it falls below any 
cut-off that one might use to define the limit of the light-front Fock space. 
That is, the dynamical colour charge decouples from the physics being beyond the cut-off and 
leaves behind a heavy constituent
quark quasiparticle relic carrying the 
quantum numbers of electric charge and isospin flavour \cite{Bass:1996iq}.
These constituent quark quasiparticles would carry colour just as a non-dynamical quantum number with colour- singlet hadrons as the physical states.
The supercritical colour Coulomb interaction is
symmetric between left- and right-handed quarks with the constituent quark 
mass term generated by the vacuum rearrangement. 
Understanding the details of such supercritical bound states and how they might enter spectroscopy are challenges for theory. 
Perhaps the full confinement potential might be a combination of usual scalar confinement and DChSB with an excited second minimum corresponding to supercritical confinement.

\section{Conclusions}

Non-perturbative glue including its topology play an important in the mass and interactions of the $\eta'$.
It is interesting to consider the possible excitation of this glue, both in the interactions of the $\eta'$ as well as in the structure of the nucleon and its resonance excitations.
Excitations of the glue generating the $\eta'$'s gluonic mass term 
as well as the possibility of an
extra minima in the confinement potential 
corresponding to novel analytic structure in the quark propagators might induce interesting phenomenology in the nucleon resonance spectrum.
Resolving 
the details poses  interesting challenges for both experiment and theory.

\section*{Acknowledgments}

I thank V. Metag for helpful discussions on aspects of $\eta'$ physics.


\begin{thebibliography}{0}


\bibitem{Shore:1998dm}
G.~M.~Shore, 
hep-ph/9812354.

\bibitem{Shore:2007yn}
G.~M.~Shore,
{\it 
Lect. Notes Phys.}  \textbf{737}, 235 (2008)

\bibitem{Witten:1979vv}
E.~Witten, 
{\it Nucl. Phys. B} \textbf{156}, 269 (1979)

\bibitem{Veneziano:1979ec}
G.~Veneziano, 
{\it Nucl. Phys. B} \textbf{159}, 213 (1979)

\bibitem{Cichy:2015jra}
K.~Cichy \textit{et al.} [ETM], 
{\it JHEP} \textbf{09}, 020 (2015)

\bibitem{DiVecchia:1980yfw}
P.~Di Vecchia and G.~Veneziano, 
{\it Nucl. Phys. B}  \textbf{171}, 253 (1980)

\bibitem{Witten:1980sp}
E.~Witten, 
{\it Annals Phys.} \textbf{128}, 363 (1980)

\bibitem{Leutwyler:1997yr}
H.~Leutwyler,
%
{\it Nucl. Phys. B Proc. Suppl.} \textbf{64}, 223 (1998)

\bibitem{Bass:1999is}
S.~D.~Bass,
%``Gluons and the eta-prime nucleon coupling constant,''
{\it Phys. Lett. B} \textbf{463}, 286 (1999) 

\bibitem{Shore:1991dv}
G.~M.~Shore and G.~Veneziano,
%``The U(1) Goldberger-Treiman relation and the proton 'spin': A Renormalization group analysis,''
{\it Nucl. Phys. B} \textbf{381}, 23 (1992)

\bibitem{Bass:2024sgg}
S.~D.~Bass,
%`
{\it Int. J. Mod. Phys. A} \textbf{39}, 2441008 (2024)

\bibitem{Bass:2004xa}
S.~D.~Bass,
{\it Rev. Mod. Phys.} \textbf{77}, 1257 (2005)

\bibitem{Aidala:2012mv}
C.~A.~Aidala, S.~D.~Bass, D.~Hasch and G.~K.~Mallot,
{\it Rev. Mod. Phys.} \textbf{85}, 655 (2013)

\bibitem{Bass:2018xmz}
S.~D.~Bass and P.~Moskal,
%``$\eta′$ and $\eta$ mesons with connection to anomalous glue,''
{\it Rev. Mod. Phys.} \textbf{91}, 015003 (2019)

\bibitem{CBELSATAPS:2013waf}
M.~Nanova \textit{et al.} [CBELSA/TAPS],
%``Determination of the ${\eta}$'-nucleus optical potential,''
{\it Phys. Lett. B} \textbf{727}, 417 (2013)

\bibitem{Metag:2017yuh}
V.~Metag, M.~Nanova and E.~Y.~Paryev,
%``Meson\textendash{}nucleus potentials and the search for meson\textendash{}nucleus bound states,''
{\it Prog. Part. Nucl. Phys.}  \textbf{97}, 199 (2017)

\bibitem{Bass:2021rch}
S.~D.~Bass, V.~Metag and P.~Moskal,
%``The \ensuremath{\eta}- and \ensuremath{\eta}'-Nucleus Interactions and the Search for \ensuremath{\eta}, \ensuremath{\eta}'- Mesic States,''
%
in I. Tanihata, H. Toki, T. Kajino (eds) 
{\it 
Handbook of Nuclear Physics}  
(Springer, Singapore, 2022), 
%doi:10.1007/978-981-15-8818-1\_39-1, 
arXiv:2111.01388 [hep-ph].

\bibitem{Bass:2005hn}
S.~D.~Bass and A.~W.~Thomas,
%``eta bound states in nuclei: A Probe of flavor-singlet dynamics,''
{\it Phys. Lett. B} \textbf{634}, 368 (2006)

\bibitem{Bass:2013nya}
S.~D.~Bass and A.~W.~Thomas,
%``QCD Symmetries in $\eta $- and $\eta '$-Mesic Nuclei,''
{\it Acta Phys. Polon. B} \textbf{45}, 627 (2014)

\bibitem{Czerwinski:2014yot}
E.~Czerwinski
%, P.~Moskal, M.~Silarski, S.~D.~Bass, D.~Grzonka, B.~Kamys, A.~Khoukaz, J.~Klaja, W.~Krzemien and W.~Oelert, 
\textit{et al.}
%``Determination of the eta'-proton scattering length in free space,''
{\it Phys. Rev. Lett.} \textbf{113}, 062004 (2014)

\bibitem{Close:1979bt}
F.~E.~Close,
{\it An Introduction to Quarks and Partons} (Academic N.Y., 1979)

\bibitem{Burkert:2022adb}
V.~Burkert, E.~Klempt and U.~Thoma,
%``Light-quark baryons,''
[arXiv:2211.12906 [hep-ph]].

\bibitem{Anisovich:2011ye}
A.~V.~Anisovich {\it et al.},
%, E.~Klempt, V.~A.~Nikonov, A.~V.~Sarantsev and U.~Thoma,
%``Nucleon resonances in the fourth resonance region,''
{\it Eur. Phys. J. A} \textbf{47}, 153 (2011)

\bibitem{Leinweber:2024psf}
D.~B.~Leinweber {\it et al.}, 
%C.~D.~Abell, L.~C.~Hockley, W.~Kamleh, Z.~W.~Liu, F.~M.~Stokes, A.~W.~Thomas and J.~J.~Wu,
%``Understanding the nature of baryon resonances,''
{\it Nuovo Cim. C} \textbf{47}, 146 (2024)

\bibitem{Anisovich:2018yoo}
A.~V.~Anisovich {\it et al.}, 
%V.~Burkert, M.~Dugger, E.~Klempt, V.~A.~Nikonov, B.~G.~Ritchie, A.~V.~Sarantsev and U.~Thoma,
%``Proton-$\eta^\prime$ interactions at threshold,''
{\it Phys. Lett. B} \textbf{785}, 626 (2018)

\bibitem{Tiator:2018heh}
L.~Tiator {\it et al.}, 
%, M.~Gorchtein, V.~L.~Kashevarov, K.~Nikonov, M.~Ostrick, M.~Had\v{z}imehmedovi\'c, R.~Omerovi\'c, H.~Osmanovi\'c, J.~Stahov and A.~\v{S}varc,
%``Eta and Etaprime Photoproduction on the Nucleon with the Isobar Model EtaMAID2018,''
{\it Eur. Phys. J. A} \textbf{54}, 210 (2018).

\bibitem{tHooft:1973alw}
G.~'t Hooft,
%``A Planar Diagram Theory for Strong Interactions,''
{\it Nucl. Phys. B} \textbf{72}, 461 (1974)

\bibitem{Witten:1979kh}
E.~Witten,
%``Baryons in the 1/n Expansion,''
{\it Nucl. Phys. B} \textbf{160}, 57 (1979)

\bibitem{Thiel:2022xtb}
A.~Thiel, F.~Afzal and Y.~Wunderlich,
%``Light Baryon Spectroscopy,''
{\it Prog. Part. Nucl. Phys.} \textbf{125}, 103949 (2022)

\bibitem{LeviSandri:2014uhc}
P.~Levi Sandri
%, G.~Mandaglio, V.~De Leo, O.~Bartalini, V.~Bellini, J.~P.~Bocquet, M.~Capogni, F.~Curciarello, A.~D'Angelo and J.~P.~Didelez, 
\textit{et al.} [GRAAL], 
%``First Measurement of the $\Sigma$ Beam Asymmetry in $\eta^{\prime}$ Photoproduction off the Proton near Threshold,''
{\it Eur. Phys. J. A} \textbf{51}, 77 (2015)

\bibitem{A2:2017gwp}
V.~L.~Kashevarov \textit{et al.} [A2],
%``Study of \ensuremath{\eta} and \ensuremath{\eta}' Photoproduction at MAMI,''
{\it Phys. Rev. Lett.} \textbf{118}, 212001 (2017)

\bibitem{Jaffe:2006jy}
R.~L.~Jaffe, D.~Pirjol and A.~Scardicchio,
%``Parity doubling among the baryons,''
{\it Phys. Rept.} \textbf{435}, 157 (2006)

\bibitem{Glozman:1999tk}
L.~Y.~Glozman,
%``Parity doublets and chiral symmetry restoration in baryon spectrum,''
{\it Phys. Lett. B} \textbf{475}, 329 (2000)

\bibitem{Brodsky:2014yha}
S.~J.~Brodsky, G.~F.~de Teramond, H.~G.~Dosch and J.~Erlich,
%``Light-Front Holographic QCD and Emerging Confinement,''
{\it Phys. Rept.} \textbf{584}, 1 (2015)

\bibitem{Bass:2021acr}
S.~D.~Bass, A.~De Roeck and M.~Kado,
%``The Higgs boson implications and prospects for future discoveries,''
{\it Nature Rev. Phys.} \textbf{3}, 608 (2021)

\bibitem{Gribov:1986vp}
V.~N.~Gribov,
%``A New Hypothesis on the Nature of Quark and Gluon Confinement,''
{\it Phys. Scripta T} \textbf{15}, 164 (1987)

\bibitem{Gribov:1999ui}
V.~N.~Gribov,
%``The Theory of quark confinement,''
{\it Eur. Phys. J. C} \textbf{10}, 91 (1999)

\bibitem{Bass:1996iq}
S.~D.~Bass and D.~Schütte,
%``The Role of the pion cloud in the interpretation of the valence light cone wave function of the nucleon,''
{\it Z. Phys. A} \textbf{357}, 85 (1997)

\end{thebibliography}
\end{document}